# FORMATION OF EMBEDDED MICROSTRUCTURES BY THERMAL ACTIVATED SOLVENT BONDING

S.H. Ng[1], R.T. Tjeung[1], Z.F. Wang[1], A.C.W. Lu[1], I. Rodriguez[2], N.F. de Rooij[3]

[1]Singapore Institute of Manufacturing Technology
[2]Institute of Materials Research and Engineering
[3]Institute of Microtechnology, University of Neuchatel

**ABSTRACT**

We present a thermal activated solvent bonding technique for the formation of embedded microstrucutres in polymer. It is based on the temperature dependent solubility of polymer in a liquid that is not a solvent at room temperature. With thermal activation, the liquid is transformed into a solvent of the polymer, creating a bonding capability through segmental or chain interdiffusion at the bonding interface. The technique has advantages over the more commonly used thermal bonding due to its much lower operation temperature (30 °C lower than the material's $T_g$), lower load, as well as shorter time. Lap shear test indicated bonding shear strength of up to 2.9 MPa. Leak test based on the bubble emission technique showed that the bonded microfluidic device can withstand at least 6 bars (87 psi) of internal pressure (gauge) in the microchannel. This technique can be applied to other systems of polymer and solvent.

## 1. INTRODUCTION

Microfluidic devices have gained tremendous interest in both academic and industrial research due to key advantages such as fast response times and low analyte consumption. Although conceptually trivial, one of the most challenging steps in the fabrication of a microfluidic device is the bonding of a structured substrate with a cover plate to create effectively sealed microchannels. Bonding methods for glass and silicon devices are often convoluted and involve the application of high temperatures for extended periods of time. Extensive surface cleaning and surface activation techniques are also required. In anodic bonding, high voltages are also required. The tolerance for the flatness of both bonding surfaces is small and this becomes challenging when wafer area scales up for large volume production. These direct bonding methods can damage sensitive structures and active components such as microelectrode arrays, waveguides and sensors. Hence, low temperature bonding methods for these materials have been developed such as adhesive bonding [1,2].

Polymer based microfluidic devices have generated a lot interest in the academic community because it is low-cost, disposable, and suitable for mass production. While techniques for the bonding of two flat surfaces of polymers have been established—e.g. adhesive bonding and thermal bonding—the bonding of two structured surfaces of polymers has its issues. With standard thermal bonding process, the microstructures deform easily, clogging the microchannels because of the high temperatures and pressures required for bonding to occur. Zhu et al. [3] recommended a temperature of 91 to 95 °C when thermal bonding polymethyl methacrylate (PMMA), as higher temperatures would result in the deformation and collapse of the microchannel. Direct thermal bonding of polymers is driven by bonding pressure (forced flow), temperature and time. However, the deformation of the microchannel is also driven by the same three factors. The best results are normally achieved at lower bonding pressures and temperatures but at a huge sacrifice of a long bonding time. In view of the limitations of thermal bonding, plasma or X-ray assisted thermal bonding processes have been developed [4,5]. In the case of the plasma assisted process, both surfaces of the polymer e.g. PMMA, are activated by plasma and at the same time becomes more hydrophilic. Thermal bonding can then be achieved with lower temperature, lower load, and moderate time, reducing the risks of microstructure deformation or clogging of microchannels. The bonding has to be carried out immediately after irradiation as the surface properties of the polymer changes quickly with time and is very sensitive to the humidity of the environment. Most direct bonding method such as the thermal bonding requires a vacuum system in order to prevent the formation of trapped air bubbles during the bonding process. Trapped air bubbles are undesirable as they would lower the bonding strength of the interface; and they exist as an inhomogeneous mismatch of materials and parameters across the bonding interface. Bubbles that exist across microstructures will cause failure of the whole device. The problem escalates when large area bonding is employed to scale up for high volume production.

Another popular method is the use of adhesives to bond two surfaces of polymers together [3,6]. However, adhesive bonding methods introduce another material to the interface which can cause compatibility issues with the fluid flowing through the microchannels. There will be a step change in material parameters across the bonding interface such as thermal properties and optical properties. Thermal mismatch can result in delamination at the interface. In many microfluidic applications, the observation and sensing technique are optical-based such as fluorescence microscopy, surface plasmon resonance and particle image velocimetry—requiring optical transparency and consistency in the materials.

Some researchers have looked at solvent bonding since it gives relatively strong bonding, without introducing a foreign adhesive material. However, adaptations have to be in place since the solvent can also destroy the microchannels. Shah et al. [7] attempted to bond two pieces of PMMA together by pumping acetone through the microchannel, and relying on capillary action to draw some of the acetone into the bonding interface. It was reported that a drop of solvent left in the microchannel for longer than 2–3 seconds would deform the microchannel. An additional step of applying acetone to the edges had to be done to ensure complete sealing. Kelly et al. [8] used melted paraffin wax as a sacrificial material to fill up the microchannels, before solvent bonding two PMMA layers together using acetonitrile. The paraffin wax was then removed by heating the bonded device and suctioning out the melted wax, followed by soaking the microchannels in cyclohexane. The whole process involved many steps but could produce higher yield than thermal bonding methods. There were also some issues with the contraction of paraffin wax upon solidification and excess solvent dissolving other parts of the device.

In this research, we look at the formation of embedded microchannels out of PMMA through a thermal activated solvent bonding process. The mechanism for solvent bonding is through the dissolution of the polymer from both bonding surfaces followed by the interdiffusion of polymer chains. The technique described here uses a liquid that is not a solvent of PMMA at room temperature but becomes so only at elevated temperatures.

## 2. SOLUBILITY OF PMMA

Polymethyl methacrylate is the synthetic polymer of methyl methacrylate. It is an amorphous thermoplastic with a density higher than water. If the chemical structure such as polarity of a polymer and a solvent molecule are alike, dissolution will occur. Hence, PMMA will not dissolve or swell in polar liquids like water and alcohols, but will dissolve in alkanes. One approach to the estimation of mutual solubility between a polymer and a solvent is to look at their solubility parameters. The Hildebrand solubility parameter, $\delta$, is the square root of the cohesive energy density, $CED$:

$$\delta = (CED)^{1/2} = \left(\frac{\Delta E_V}{V}\right)^{1/2}$$

where $\Delta E_V$ is the cohesive energy (or energy of vaporization) and $V$ is the molar volume. The cohesive energy represents the energy required to break all cohesive bonds to convert a liquid to a gas. Major cohesive interactions existing in organic materials are van der waals forces, permanent dipole interactions and hydrogen bonding. The solubility parameter approach is based on the enthalpy of the interaction between the solvent and polymer. With the basic principle of "like dissolves like," liquids with similar solubility parameters will be miscible, while polymers will be soluble in liquids that have solubility parameters not too different from theirs. PMMA has a solubility parameter of 20.18 $(MJ/m^3)^{1/2}$ [9]. The polymer is not soluble is water ( 47.9 $(MJ/m^3)^{1/2}$ ) but dissolves readily in acetone ( 19.9 $(MJ/m^3)^{1/2}$ ) and dichloromethane ( 20.3 $(MJ/m^3)^{1/2}$ ). PMMA, however, does not dissolve easily in isopropanol ( 23.5 $(MJ/m^3)^{1/2}$ ) at room temperature.

Thermodynamically, for a polymer to dissolve in a liquid spontaneously, the free energy of mixing, $\Delta G_M$, must be less than or equal to zero [10]:

$$\Delta G_M = \Delta H_M - T\Delta S_M$$

where $\Delta H_M$ is the heat of mixing, $T$ is the absolute temperature and $\Delta S_M$ is the entropy change in the mixing process. Hence, increasing temperature has the effect of lowering the free energy of mixing, thereby promoting dissolution. While the solubility parameter of polymers does not change much with temperature, the solubility parameter for liquids does. The cohesive energy is also related to the absolute temperature by the following relation:

$$\Delta E_V = \Delta H_V - RT$$

where $\Delta H_V$ is the molar heat of vaporization and R is the gas constant. Hence, an increase in the temperature will lead to a decrease in the solubility parameter of liquids. A liquid with a higher solubility parameter than the polymer might not dissolve the polymer at room temperature; but with a temperature increase, its solubility parameter will drop turning it into a solvent for that polymer. In some cases, subsequent temperature increase might turn the liquid back to a non solvent, as its solubility parameter drops way below that of the polymer.

We conduct lap shear tests to look at the bond strength of blank PMMA samples using the thermal activated thermal bonding. Base on the results, the process window is narrowed down for bonding experiments on samples with microchannels. Leak tests based on the bubble emission technique are also conducted on the bonded samples.

### 3. LAP SHEAR TESTS

There are three main methods for testing plastic bonding: the tensile test involving butt joints, the shear test involving lap joints and the peel test for peel joints. Other tests consist of the cantilever beam test, blister test and cone test [11]. The standard tensile test usually involves bars or rods bonded end to end forming the butt joint. Hence, performing tensile testing on thin sheets can be challenging since mounting to the holder is critical in preventing modifications to the existing bond. Epoxy is normally used for mounting and the most common occurrence is failure at the mounting (interface between the epoxy and the plastic sheet)—especially those involving high strength bonding techniques like solvent bonding. The peel test usually requires at least one of the substrate to be flexible.

The lap shear test is a standard test method for determining the shear bond strength of adhesively bonded plastics. It is especially useful when thin sheets of plastics are bonded together. Hence, the method is adopted in this research. In each specimen, two strips of PMMA are bonded together with an overlapping bond area (see Fig. 1(a)) in accordance to the lap shear test (ASTM D1002-05). The PMMA used in all experiments are Poly-A cast, acrylic sheets purchased from Dama Enterprise (Singapore). The dimensions of each strip of PMMA before bonding are: Width (25.4 mm), Length (101.6 mm), Overlap (12.7 mm x 25.4 mm), Grip area (12.7 mm x 25.4 mm). The bonded specimens are subjected to a pull test to determine the load at failure. An Instron 4505 Mechanical Tester is used to perform the pull test (see Fig. 1(c)). The cross head speed is set at 0.5 mm/min. Figure 1(b) shows a typical loading cycle where the pull strength increased with the elongation of the specimen until failure where the force dropped to zero suddenly. The bonding of each test specimen is done on top of a hot plate with set temperature. Dead weights are placed on top of the lap joint after a few drops of isopropanol (IPA) have been deposited. The IPA used in this research is 99.9% pure technical grade supplied by Sino Chemical Co Pte Ltd (Singapore). Air bubbles can be easily avoided by careful placement of the PMMA strips. Excess IPA is squeezed out from the bonding interface leaving a uniform thin film when the load is applied.

The range of the bonding parameters studied is: Load (1 – 5 kg), Temperature (25-80 °C), Time (5 – 15 min.). Three repeats were performed for each set of parameters and their averages taken and plotted. A calibration is conducted to get the actual temperature at the bonding interface. A fine thermocouple (76 μm diameter wires) is inserted into interface between two PMMA substrates while another thermocouple is placed on the hotplate surface. The calibration curve is shown in Fig. 2. In the hotplate temperature range from 70 to 80 °C, the typical temperature drop over the PMMA substrate is about 10 °C.

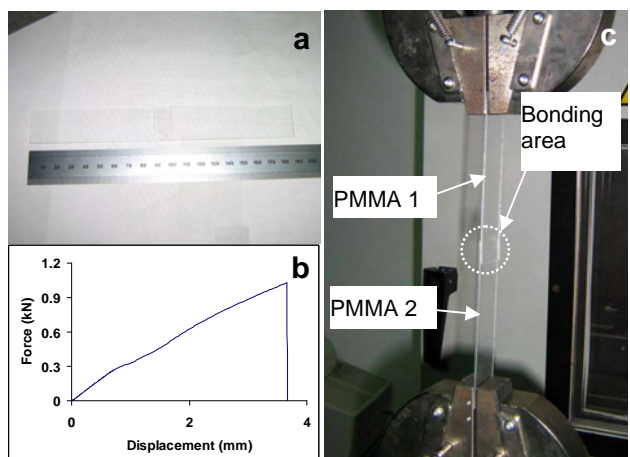

Fig. 1 Lap shear test specimen (a), Typical loading cycle (b), Lap shear test setup (c).

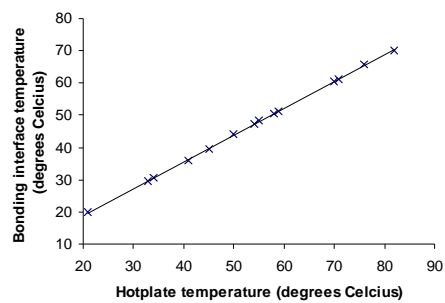

Fig. 2 Calibration of bonding interface and hotplate temperatures.

### 4. LEAK TESTS

Further experiments are performed to investigate the bonding performance of substrates with microchannels. The aim is to test liquid flow in the microchannels at atmospheric pressure, as well as the pressure limit of microfluidic devices bonded with these parameters. PMMA samples with laser cut channels are utilized for the experiments. Each sample dimensions are 12.7 mm x 25.4 mm—same as the bond area in the lap shear test specimens. A 15 mm long microchannel is created on the surface of the PMMA sample by excimer laser. Two 1 mm diameter holes were drilled on both ends of the microchannel. Figure 3 shows the SEM images of the laser cut microchannel. The microchannel is 85 µm deep and 150 µm wide. A set of bonding experiments are performed on the laser cut samples, filtering off conditions (< 1 kg, <50 °C ) that do not result in bonding as indicated in the earlier lap shear tests. The laser cut and drilled samples are bonded to another piece of PMMA of similar size using the same technique described earlier.

In the literature, researchers have used different leak tests to measure the performance of their bonded devices. The most commonly used method is passing a fluorescence liquid or a dye [3] through the microchannels—looking out for any leak assisted or unassisted optically. The test gives fast results as to whether the device is leaking or blocked. The internal pressure it is subjected to is typically slightly above atmospheric. In pressure limit tests, there are some variants giving different range of results. Most of these tests involve generating an elevated internal gas pressure within the microchannels by connecting the inlet of the device to a compressed gas supply or gas pump, while plugging all other outlets. In the bubble emission test [12], the device is immersed in water and the gas pressure increased until a bubble is seen, indicating a leak has occurred. Another test involved pressuring the microchannel with a liquid or gas and monitoring the pressure decay with time [13]. One test involved pressurizing the device until the two bonded layers of substrates separate [7,8]. This test typically gives very high pressure limits due to the more severe failure criterion.

In this research, we conduct a liquid flow test at atmospheric pressure, followed by a pressure limit test by the bubble emission technique. A flow test at atmospheric pressure is carried out on each bonded sample by passing iodine solution through the microchannels. Leaks are detected with the help of a microscope and samples with clogged microchannels identified. Samples that passed the flow test are subjected to an elevated pressure test to determine their pressure limits. The test is similar to the bubble emission technique for detecting leaks (ASTM E515-05), except that the samples are subjected to increasing pressures until failure. The normal limit of sensitivity for this test method is $10^{-5}$ Std. $cm^3/s$. The bonded sample is prepared by sealing one end of the microchannel with epoxy and connecting the other end to a compressed dry air supply via tubing. It is then submerged under water in a glass beaker, making sure that no air bubble is sticking to the sample. Shown in Fig. 4, a pressure gauge is used to monitor the static pressure during the test. A magnifier with illumination is used to assist the observation of gas bubbles during a leak. The procedure is to increase the air pressure in the microchannel to 1 bar above atmospheric pressure and to hold it at that pressure for 2 minutes while observing for any bubbles developing from the sample. If no bubble is observed, the pressure is increased by another bar and held for another 2 minutes. A bubble that forms or grows would indicate a leak and the experiment stopped indicating that the sample has failed at that pressure. A drop in the displayed reading of the pressure gauge would also indicate a leak. Otherwise, the experiment would carry on with increasing pressure until the final gauge pressure of 6 bars (87 psi). This is also the maximum pressure of the compressed air supply.

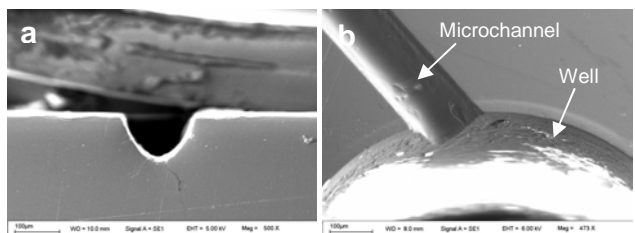

Fig. 3 Scanning electron micrographs of laser cut microchannels on PMMA.

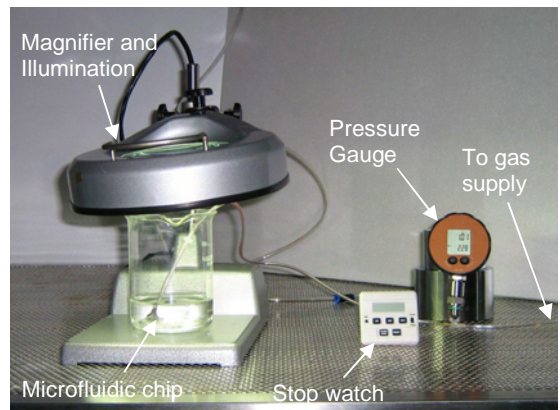

Fig. 4 Leak test setup.

## 5. RESULTS AND DISCUSSION

A set of lap shear test results can be seen in Fig. 5(a). The general trend is stronger bonding with longer time. Temperature is a sensitive parameter especially in the region from 60 °C to 70 °C. Below 60 °C, the bond strength is minimal (less than 0.1 kN except for one case). The bond strength ramps up to 0.94 kN (translates to a shear strength of 2.9 MPa) at 70 °C. Above 70 °C, the increase in bond strength is slow. In most cases (not shown), the bond strength is zero at 1 kg. The maximum shear bond strength is comparable to those reported in the literature. Brown et al. [14] reported maximum bond strength of 5.5 MPa using solvent bonding—five to ten times greater than the thermally bonded PMMA devices—based on the results of a similar lap shear test.

At room temperature, no bonding occurs in all cases. In fact, there is no bonding even after an extended period of time where all the IPA has evaporated from the interface. This phenomenon can be explained by the temperature dependence solubility of PMMA in IPA. At lower temperatures, PMMA is not soluble in IPA. But as temperature increases, it gets more and more soluble as their solubility parameters approach each other. Interdiffusion [15,16] of PMMA molecules from each surface increases as a result of increased polymer segment or chain mobility. The strength of the bonding is increased due to the increased interdiffusion. The strength of the bond depends on the degree of chain entanglement and the thickness of the diffuse interface. In solvent bonding, longer range segmental or molecular movement can occur than in thermal bonding. Hence, solvent bonding is a faster process (up to a few minutes) [7] while thermal bonding can take as long as a few hours [3,17]. In addition, the bond strength by solvent bonding is generally much higher than by thermal bonding [8,14].

The bond strength when using 1 kg load is low in most cases and should theoretically be at zero at some threshold load between 0 to 1 kg. As seen in Fig. 5(b), the bond strength increases rapidly as the load increases from 1 kg. The bond strength peaks in the range of 0.8 to 1 kN under loads of 3 to 5 kg. The general observation is higher bond strength at longer times and higher temperatures. Since diffusion is known to be independent of contact pressure, the phenomenon could be as a result of increased forced flow. At high loads of 10 and 15 kg, the bond strength is actually decreased. This is in contrast to the phenomenon seen in "dry" direct thermal bonding where an increase in bonding pressure led to an increase in bonding strength [3]. One reason could be the "squeeze film" effect, where the thickness of the thin film of solvent existing at the bonding interface is dependent on the load across it. In the Greenwood and Williamson model [18] for the contact of two surfaces, the load is supported by the asperities due to the roughness and waviness of the surfaces. As load is increased, the asperities deformed according to the Hertzian contact model leading to the decrease in the separation distance between the two surfaces. Consequently, there is a decrease in the amount of solvent at the bonding interface as more got "squeezed" out at higher loads. The starvation of solvent available for bonding is believed to lead to a decrease in the bond strength.

The flow test at atmospheric pressure indicates that all samples pass the test except for those bonded at 80 °C. These samples that passed the flow test are subjected to the pressure limit test. Samples bonded at 60 °C and below failed at a gauge pressure of less than 1 bar, while samples bonded at 70 °C withstood gauge pressures of up to 6 bars (87 psi). Figure 6 shows a scanning electron micrograph of the cross section of a bonded sample. Shah et al. [7] reported that their PMMA device created by solvent bonding could withstand 80 psi internal pressure while Kelly et al. [8] reported 2250 psi. Both tested their devices using compressed gas until the bonded layers separated. The shape and size of the microchannel is preserved when compared with that shown in Fig. 3.

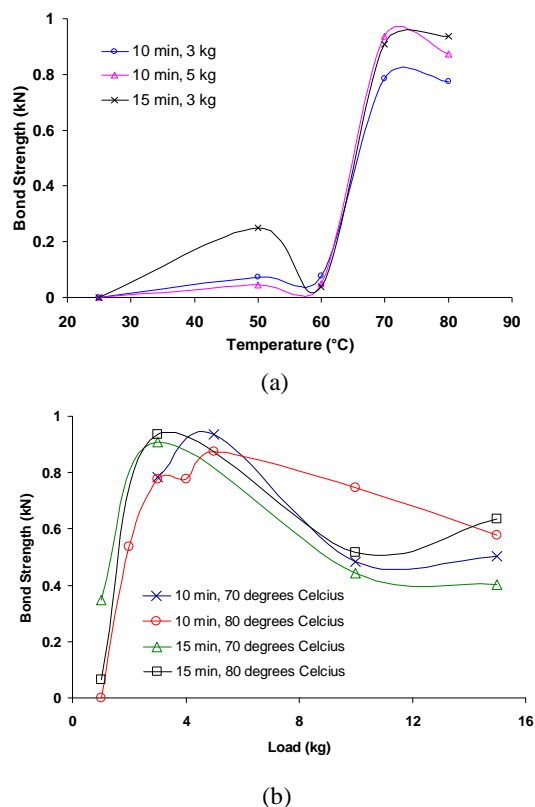

(a)

(b)

Fig. 5 Results of lap shear test.

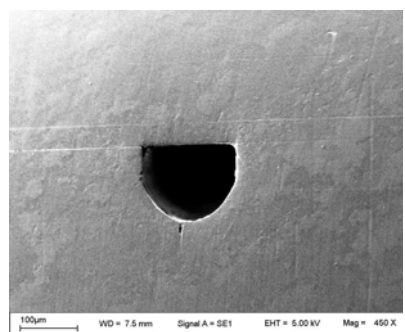

Fig. 6 Scanning electron micrograph showing the cross section of a bonded microchannel.

## 6. CONCLUSIONS

While using a thermally activated solvent has the effect of softening the polymer (which has similar effect to heating it to a much higher temperature), it involves only the very top layer of the polymer that is in contact with the solvent. The "dry" heating method involves softening the bulk of the polymer. It is possible, by controlling the process parameters (temperature and time), to limit the diffusion and softening of the polymer to certain depths using the thermally activated solvent method. This will create a surface layer suitable for bonding, while having a structurally sound bulk material to reduce deformation. The nature of the method allows time and flexibility for alignment of the substrates for bonding (especially multilayer devices) since the solvent is only activated at elevated temperatures. Flooding the interface with the solvent during assembly of the layers helps to remove air bubbles. Excess solvent is squeezed out when the load is applied creating a thin uniform film of solvent. Evaporation of the solvent at room temperature is slow once the solvent is in the bonding interface, allowing ample time for alignment and adjustment. The solvent strength can be tuned during the bonding process by temperature control. This technique can be applied to other systems of polymer and solvent.

## ACKNOWLEDGEMENT

This research is funded by the Agency for Science, Technology and Research (A*STAR), Singapore.